\documentclass[aps,prl,twocolumn]{revtex4}
\newcommand {\be}{\begin{equation}}
\newcommand {\ee}{\end{equation}}
\newcommand {\bey}{\begin{eqnarray}}
\newcommand {\eey}{\end{eqnarray}}

\usepackage{epsfig}
\usepackage{amsfonts}
\usepackage{amsmath}
\usepackage{mathbbol}

\usepackage{hyperref}

\begin{document}

\title{Communication Complexity of Channels in General Probabilistic Theories}

\author{A. Montina, M. Pfaffhauser, S. Wolf}
\affiliation{Facolt\`a di Informatica, 
Universit\`a della Svizzera Italiana, Via G. Buffi 13, 6900 Lugano, Switzerland}

\date{\today}

\begin{abstract}
The communication complexity of a quantum channel is the minimal amount of classical
communication required for classically simulating the process of preparation, transmission
through the channel, and subsequent measurement of a quantum state. At present, only 
little is known about this quantity. 
In this paper, we present a procedure for systematically evaluating the communication 
complexity of channels in any general probabilistic theory, in particular quantum theory.
The procedure is constructive and provides the most efficient classical protocols.
We illustrate this procedure by evaluating the communication complexity of a 
quantum depolarizing channel with some finite sets of quantum states and measurements.
\end{abstract}

\maketitle

Quantum communication has proved to be much more powerful than its classical counterpart.
Indeed, quantum channels can provide an exponential saving of communication resources 
in some distributed computing problems~\cite{buhrman}, where the task is to evaluate 
a function of data held by two or more parties. A natural measure of power of quantum
communication in a two-party scenario is provided by the communication complexity of a
quantum channel, which is defined as the minimal amount of classical communication
required for classically simulating the process of preparation, transmission through
the channel, and subsequent measurement of a quantum state. Indeed, it is clear that a
quantum channel cannot replace an amount of classical communication greater than its 
communication complexity. Thus, this quantity sets an ultimate limit to the power of 
quantum communication in a two-party scenario in terms of classical resources.

At present, only little is known about the communication complexity of quantum channels.
Toner and Bacon proved that two classical bits are sufficient to simulate the communication
of a single qubit~\cite{toner}. In the case of parallel simulations, the communication
can be compressed so that the asymptotic cost per simulation is about $1.28$ 
bits~\cite{montina}. Simulating the communication of $n$ qubits requires an amount
of classical communication greater than or equal to $2^n-1$ bits~\cite{montina2}. 
However, no upper bound is known. 

In this paper, we present a general procedure for systematically evaluating the communication 
complexity of channels in any general probabilistic theory, in particular quantum theory. 
The procedure relies on the reverse Shannon theorem~\cite{rev_s} and a strategy discussed in 
Refs.~\cite{montina,montina3}. There, it was shown that any classical simulation protocol 
can be turned into a protocol with communication cost equal to the
classical mutual information between the quantum state and the communicated variable
of the parent protocol. A similar role of the mutual information is played in
the context of classical simulations of measurements~\cite{wilde}.
We illustrate the procedure by evaluating the communication complexity of a 
quantum channel with some finite sets of quantum states and measurements.

A protocol simulating a quantum channel actually simulates a process of preparation,
transmission through the channel and subsequent measurement of a quantum state.
For the sake of simplicity, we will focus on quantum channels, but the
following discussion can be easily generalized to any probabilistic theory,
as pointed out later.
The simulated quantum scenario is as follows. A party, say Alice, prepares $n$ qubits
in some quantum state $|\psi\rangle\langle\psi|\equiv\hat\rho$ according to an unknown 
probability distribution $\rho(\psi)$. Then, she sends the qubits to another party,
say Bob, through a quantum channel with associated superoperator $\cal L$. Finally, 
Bob generates an outcome by performing a measurement 
${\cal M}=\{\hat E_1,\hat E_2,\dots\}$, where $\hat E_i$ are positive semidefinite
self-adjoint operators labeling events of the measurement $\cal M$. 
The quantum probability of getting the $w$-th outcome $\hat E_w$, given $|\psi\rangle$ 
and $\cal M$, is
\be
P_{\cal L}(w|\psi,{\cal M})\equiv Tr\left[\hat E_w {\cal L}(\hat \rho)\right].
\ee
In a classical simulation, the quantum channel between Alice and Bob is replaced by 
classical communication. A classical protocol is as follows.
Alice sets a variable, say $k$, according to a probability distribution 
$\rho(k|y,\psi)$ that depends on the quantum state $|\psi\rangle$ and, possibly, 
a random variable $y$ shared with Bob. Thus, there is a mapping from the
quantum state to a probability distribution of $k$,
\be\label{map_psi_k}
|\psi\rangle\xrightarrow{y} \rho(k|y,\psi).
\ee
Alice sends $k$ to Bob, who simulates 
a measurement $\cal M$ by generating an outcome $\hat E_w$ with a probability
$P(w|k,y,{\cal M})$. The protocol exactly simulates the quantum channel
if the probability of $\hat E_w$ given $|\psi\rangle$ is equal to the quantum
probability, that is, if
\be
\sum_k\int dy P(w|k,y,{\cal M})\rho(k|y,\psi)\rho(y)=P_{\cal L}(w|\psi,{\cal M}),
\ee
where $\rho(y)$ is the probability density of the random variable $y$.
Let us denote by $\rho(k|y)\equiv\int d\psi\sum\rho(k|y,\psi)\rho(\psi)$
the marginal conditional probability of $k$ given $y$. As defined in Ref.~\cite{montina3},
the communication cost, say $\cal C$, of the classical simulation is the maximum,
over the space of distributions $\rho(\psi)$, of the Shannon
entropy of the distribution $\rho(k|y)$ averaged over $y$, that is,
\be
{\cal C}\equiv\max_{\rho(\psi)} H(K|Y),
\ee
where $H(K|Y)\equiv-\int dy\rho(y)\sum_k \rho(k|y)\log_2\rho(k|y)$.

Shannon's source coding theorem~\cite{cover} establishes an operational meaning of $\cal C$,
as discussed in Ref.~\cite{montina3}. Indeed, suppose 
that $N$ independent simulations of $N$ quantum channels are performed in parallel. Let $k^i$ be the 
variable prepared with probability $\rho(k^i|y,\psi^i)$, where $|\psi^i\rangle$ is the quantum
state prepared for the $i$-th quantum channel. Instead of communicating directly the variables $k^i$,
we can encode them into a global $k$, so that
the average number of communicated bits per simulation approaches $\cal C$ 
with vanishing error as $N$ goes to infinity. The quantity $\cal C$ is the minimal 
compression rate for the worst case distribution $\rho(\psi)$. Furthermore, 
it is possible to show that there is an
compression code that is optimal for the worst case and has a compression rate 
independent of the actual distribution $\rho(\psi)$ and equal to $\cal C$.

We define the {\it communication complexity} [denoted by ${\cal C}_{min}({\cal L})$] of a quantum 
channel $\cal L$ as the minimal amount of classical communication $\cal C$ required by an exact 
classical simulation of the quantum channel, given any measurement $\cal M$ (in a possible 
alternative definition, only projective measurements would be allowed).
Let ${\bf S}\equiv\{|\psi_1\rangle,\dots,|\psi_S\rangle\}$ and
${\bf M}\equiv\{{\cal M}_1,\dots,{\cal M}_{M}\}$ be a set of $S$ quantum states and 
$M$ measurements, respectively. We define the communication complexity, say 
${\cal C}_{min}({\bf G})$, of the quantum game $({\cal L},{\bf S},{\bf M})\equiv{\bf G}$ 
as the minimal amount of classical communication required to simulate the quantum channel 
$\cal L$ with the restriction that the quantum states and
the measurements are elements of $\bf S$ and $\bf M$, respectively. 
The quantities ${\cal C}_{min}({\cal L})$ and ${\cal C}_{min}({\bf G})$ 
are functionals of $\cal L$ and $\bf G$, respectively.

Let us consider the case of $N$ quantum channels. In a general
parallel simulation, the communicated variable $k$ is generated according to a
probability distribution $\rho(k|y,\psi^1,\psi^2,\dots,\psi^N)$ depending on the whole
set of $N$ prepared quantum states $|\psi^1\rangle,\dots,|\psi^N\rangle$. Thus, the
single-shot map~(\ref{map_psi_k}) is replaced by
\be\label{prob_paral}
\{|\psi^1\rangle,\dots,|\psi^N\rangle\}\xrightarrow{y}\rho(k|y,\psi^1,\psi^2,\dots,\psi^N).
\ee
The asymptotic communication cost, say ${\cal C}^{asym}$, is equal to 
$\lim_{N\rightarrow\infty}{\cal C}^{par}/N$, ${\cal C}^{par}$ being the cost of the 
parallelized simulation. The definition of ${\cal C}^{par}$ is similar to that of $\cal C$, 
with the difference that the maximization is made over the space of the distributions
$\rho(\psi^1,\dots,\psi^N)$.
We define the asymptotic communication complexity, ${\cal C}_{min}^{asym}({\cal L})$, of a quantum
channel $\cal L$ as the minimal asymptotic communication cost required for simulating the 
channel. The asymptotic communication complexity ${\cal C}_{min}^{asym}({\bf G})$
of the game ${\bf G}$ is similarly defined.

Given a game ${\bf G}=({\cal L},{\bf S},{\bf M})$,
let ${\bf w}=\{w_1,\dots,w_M\}$ be an $M$-dimensional array whose $m$-th element is one of
the possible outcomes of the $m$-th measurement ${\cal M}_m\in{\bf M}$. We denote by 
$s=1,\dots,S$ and $m=1,\dots,M$ discrete indices labelling the elements of $\bf S$ and $\bf M$, 
respectively. The summation over every index in $\bf w$ but the $m$-th one, which is set equal to 
$w$, is concisely written as follows,
\be
\sum_{w_1,\dots,w_{m-1},w_m=w,\dots,w_M} \rightarrow\sum_{{\bf w},w_m=w}
\ee
\newline
{\bf Definition.} Given a game ${\bf G}=({\cal L},{\bf S},{\bf M})$,
the set ${\cal V}({\bf G})$ contains any conditional probability $\rho({\bf w}|s)$ 
over the sequence ${\bf w}=\{w_1,\dots,w_M\}$
whose marginal distribution of the $m$-th variable is the quantum distribution of the
outcome $w_m$ given the quantum state $s$ and the measurement $m$, for any $s$, $m$. 
In other words, the set ${\cal V}({\bf G})$ contains any $\rho({\bf w}|s)$ satisfying the 
constraints
\be\label{constraints}
\sum_{{\bf w},w_m=w} \rho({\bf w}|s)=P_{\bf G}(w|s,m), \forall s,m \text{ and } w,
\ee
where
$$P_{\bf G}(w|s,m)\equiv P_{\cal L}(w|\psi_s,{\cal M}_m)$$
is the quantum probability 
of getting the $w$-th outcome of the measurement ${\cal M}_m$ 
given the quantum state $|\psi_s\rangle$. 
\vspace{2mm}
\newline
The set ${\cal V}({\bf G})$ is surely non-empty. A function in ${\cal V}({\bf G})$ is 
$\rho({\bf w}|s)=P_{\bf G}(w_1|s,1)\times \dots\times P_{\bf G}(w_M|s,M)$, where the 
variables $w_1,\dots,w_M$ are uncorrelated.
The definition of ${\cal V}({\bf G})$ can be easily extended to any general probabilistic 
theory, where $P_{\bf G}(w|s,m)$ is replaced by different conditional probabilities. 
For the sake of concreteness, we will refer to the quantum case, but the following
discussion does not rely on any precise form of $P_{\bf G}(w|s,m)$ and applies to more general
theories.

A pivotal classical protocol for the quantum game $\bf G$ is as follows.
\vspace{2mm}
\newline
{\bf Master protocol.} 
Alice generates the array $\bf w$ according to a conditional probability $\rho({\bf w}|s)\in
{\cal V}({\bf G})$.
Then, she sends $\bf w$ to Bob. Bob simulates the measurement ${\cal M}_m$ by outputting
the outcome $w_m$.
\vspace{2mm}
\newline
The definition of ${\cal V}({\bf G})$ implies that this protocol exactly simulates the quantum game 
$\bf G$. A classical channel from a variable $x_1$ to $x_2$ is defined by the conditional 
probability of getting $x_2$ given $x_1$. Its capacity is the maximum of the mutual information 
between $x_1$ and $x_2$ over the space of probability distributions $\rho(x_1)$~\cite{cover}. 
Using the strategy discussed in Ref.~\cite{montina} and the reverse Shannon theorem~\cite{rev_s},
it is possible to prove that a master protocol can be turned into a child protocol for
parallel simulations whose asymptotic communication cost is equal to the capacity of the 
classical channel $\rho({\bf w}|s)$.
\vspace{2mm}
\newline
{\bf Lemma 1.} Given a conditional probability $\rho({\bf w}|s)\in{\cal V}({\bf G})$, there is a 
child protocol, simulating in parallel $N$ quantum games $\bf G$, whose asymptotic communication 
cost per game is equal to the capacity of the channel $\rho({\bf w}|s)$ as $N$ goes to infinity.
\vspace{2mm}
\newline
{\it Proof.} In a parallel simulation of $N$ games $\bf G$ through $N$ master protocols,
Alice sends an array $\bf w$ to Bob for each game. This array is generated
with probability $\rho({\bf w}|s)$. Let $C({\bf W}|S)$ be the capacity of the channel 
$T:s\rightarrow{\bf w}$. The child protocol is as follows. Instead of sending ${\bf w}$, 
Alice sends an amount of information, say ${\cal C}(N)$, that allows Bob to 
generate $\bf w$ for every game $\bf G$ according to the probability $\rho({\bf w}|s)$. The 
reverse Shannon theorem states that this can be accomplished with a cost ${\cal C}(N)$ such that 
$\lim_{N\rightarrow\infty}{\cal C}(N)/N=C({\bf W}|S)$,
provided that the receiver and sender share some random variable. 
$\square$
\newline
A constructive proof of the reverse Shannon theorem and its one-shot version were provided in 
Ref.~\cite{harsha}. This gives an explicit procedure for deriving the child protocol associated with 
$\rho({\bf w}|s)$.

The first main result is the following theorem about the asymptotic communication complexity.
Later on, we will consider the single-shot case.
\vspace{2mm}
\newline
{\bf Theorem 1.} 
The asymptotic communication complexity of the game ${\bf G}=({\cal L},{\bf S},{\bf M})$ is the 
minimum of the capacity of the classical channels $\rho({\bf w}|s)$ in the set ${\cal V}({\bf G})$.
\vspace{1mm}
\newline
Theorem~1 states that the asymptotic communication complexity of the
game ${\bf G}$ is equal to the quantity
\be\label{cal_D}
{\cal D}({\bf G})\equiv\min_{\rho({\bf w}|s)\in {\cal V}({\bf G})}\left(\max_{\rho(s)} I({\bf W};S)\right),
\ee
where $I({\bf W};S)$ is the mutual information between the stochastic variables $\bf w$ and $s$.
This theorem and Lemma~1 provide a constructive method for deriving the best
protocol with communication cost equal to ${\cal D}({\bf G})$.
It is sufficient to evaluate the conditional probability $\rho({\bf w}|s)$ that solves
the minimax problem stated in Eq.~(\ref{cal_D}) and to use the procedure in 
Ref.~\cite{harsha} for deriving the associated child protocol. The proof of the theorem
is provided in the appendix. It relies on Lemma~1 and the 
data-processing inequality~\cite{cover}. Lemma~1 implies that there is a
protocol whose communication cost is ${\cal D}({\bf G})$, that is, 
${\cal C}_{min}^{asym}({\bf G})\le{\cal D}({\bf G})$. Furthermore, the communication cost of 
any simulation cannot be strictly smaller than ${\cal D}({\bf G})$. 
This is proved by showing through the data-processing inequality that any simulation protocol 
with communication cost $\cal C$ induces a master protocol with associated capacity $C({\bf W}|S)$
smaller or equal to $\cal C$.
Thus, ${\cal C}_{min}^{asym}({\bf G})={\cal D}({\bf G})$. 

Theorem~1 and Lemma~1 have their one-shot versions.
\newline
{\bf Lemma 2 (One-shot version of Lemma~1).} Given a conditional probability $\rho({\bf w}|s)
\in{\cal V}({\bf G})$, there is protocol simulating a quantum game $\bf G$ such that
\be\nonumber
C_{ch}\le {\cal C}\le C_{ch}+2\log_2(C_{ch}+1)+2\log_2 e,
\ee
where ${\cal C}$ and $C_{ch}$ are the communication cost of the simulation and the
capacity of the channel $\rho({\bf w}|s)$.
\vspace{2mm}
\newline
The proof is similar to that of Lemma~1 and relies on the one-shot version of the reverse Shannon
theorem proved in Ref.~\cite{harsha}.
\vspace{2mm}
\newline
{\bf Theorem 2 (One-shot version of Theorem~1).} The communication complexity ${\cal C}_{min}({\bf G})$
of the game $\bf G$ satisfies the inequalities 
\be\nonumber
{\cal D}({\bf G})\le {\cal C}_{min}({\bf G})\le {\cal D}({\bf G})+2\log_2[{\cal D}({\bf G})+1]+2\log_2 e,
\ee
where ${\cal D}({\bf G})$ is given by Eq.~(\ref{cal_D}) and it is equal to the asymptotic 
communication complexity of the game ${\bf G}$ (Theorem~1).
\vspace{2mm}
\newline
The first inequality is a trivial consequence of Theorem~1, as the asymptotic communication complexity
cannot be larger than the communication complexity. The second inequality comes from Lemma~2.

Thus, the communication complexity of a quantum channel is about equal to the asymptotic communication 
complexity, apart from a possible additional cost that does not grow more than the logarithm of the
asymptotic communication complexity. The asymptotic communication complexity of a quantum channel is 
obtained in the limit $S,M\rightarrow\infty$ with the sets ${\bf S}$ and $\bf M$ densely covering the
space of quantum states and measurements, respectively.

To illustrate these results, we have evaluated the communication complexity of the following game 
$\bf G$ for a binary quantum depolarizing channel. The channel is a map from a Bloch vector $\vec v$ 
to $\gamma\vec v$, where $0\le\gamma\le1$. The channel is noiseless or completely erasing if
$\gamma=1$ or $0$, respectively.
Let us denote by $\vec v_x$ the tridimensional vectorial function
$\left(\cos\frac{\pi x}{M},\sin\frac{\pi x}{M},0\right)$, where $x$ is a real number.
The measurements are projections in a two-dimensional Hilbert space.
The eigenvectors of the $m$-th measurement in $\bf M$ correspond to the Bloch vectors 
$\pm\vec v_m$ with $m=1,\dots,M$ and outcomes $w=\pm1$. 
The set $\bf S$ contains all the $2M$ eigenvectors, that is, $\vec v_s$ with $s=1,\dots,2M$.
the quantum probability of getting $w$ given $s$ and $m$ is
\be
P_{\bf G}(w|s,m)=\frac{1}{2}\left\{1+w \gamma\cos\left[\frac{\pi}{M}(s-m)\right]\right\}.
\ee
Since $P_{\bf G}$ is invariant under the transformation $s\rightarrow s+1$ and $m\rightarrow m+1$,
the distribution $\rho(s)$ solving the minimax problem in Eq.~(\ref{cal_D}) is, by symmetry,
uniform. Thus the minimax problem is reduced to a minimization problem.
We have evaluated algebraically the asymptotic communication complexity up to $M=4$. The
distributions $\rho({\bf w}|s)\in{\cal V}({\bf G})$ with minimal capacity for $M=2,3,4$ are summarized
by the analytical equation
\be\label{rho_min}
\rho({\bf w}|s)=\sum_{k=1}^{2M}P({\bf w}|k)\rho(k|s)
\ee
with
\bey
\label{condi_KS}
P({\bf w}|k)=\prod_{m=1}^M\theta\left(w_m\vec v_m\cdot\vec v_{k+p/2}\right), \\
\label{rho_ks}
\rho(k|s)=f(k,s)+\sqrt{\lambda+f(k,s)}
\eey
where $p=0$ $(1)$ if $M$ is odd (even),
$\lambda$ is a constant determined by the normalization $\sum_{k=1}^{2M}\rho(k|s)=1$ 
and
\be\label{f_funct}
f(k,s)\equiv \frac{\gamma}{2}\sin\left(\frac{\pi}{2M}\right)\vec v_{k+p/2} \cdot\vec v_s.
\ee
It is easy to prove that $\rho({\bf w}|s)$ is an element of ${\cal V}({\bf G})$
for any $M\ge2$.
For $\gamma=1$ (noiseless channel), these equations are the discrete version of the Kochen-Specker 
model~\cite{ks} with the constraint that the hidden variable $\vec v_{k+p/2}$ is a vector lying on a 
plane.

\begin{figure}
\epsfig{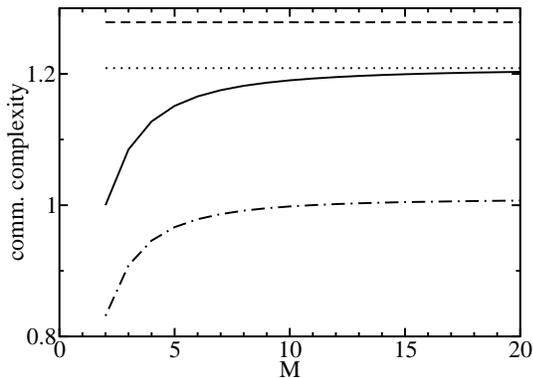}
\caption{Asymptotic communication complexity for $\gamma=1$ (solid line) and $\gamma=0.95$ 
(dashed-dot line) as a function of the number $M$ of measurements 
(the lines interpolate the data as a guide to the eyes). The dot line represents the asymptotic limit 
of the solid line for $M\rightarrow\infty$. The dashed line is the communication cost of the model 
in Ref.~\cite{montina}, working for any projective measurement on the qubit and noiseless
channel. }
\label{fig1}
\end{figure}

Since $P({\bf w}|k)$ is a noiseless channel, the capacity of $\rho({\bf w}|s)$ is equal
to the capacity of $\rho(k|s)$. Thus, the asymptotic communication complexity is
\be\label{com_complex}
{\cal C}_{min}^{asym}({\bf G})=\max_{\rho(s)} I(k;s)
\ee
In particular, for a noiseless channel
\be\label{cc_noiseless}
{\cal C}_{min}^{asym}({\bf G})={\cal N}\sum_{n=\frac{1-M}{2}}^{\frac{M-1}{2}}
\cos\left(\frac{\pi n}{M}\right)\log\left[2M{\cal N}\cos\left(\frac{\pi n}{M}\right)\right],
\ee
where ${\cal N}=\sin\left(\frac{\pi}{2M}\right)$. Note that the sum index $n$ is not an integer 
when $M$ is even. We have numerically verified the validity of the analytical equations for
$M$ up to $20$ and some values of $\gamma$. The simulations are in agreement with
Eqs.~(\ref{rho_ks},\ref{f_funct},\ref{com_complex}) within the machine precision.
The numerical data for $\gamma=1$ and $0.95$ are reported in Fig.~\ref{fig1}. 

If we extrapolate Eq.~(\ref{cc_noiseless}) to arbitrary $M$, we have 
$
\lim_{M\rightarrow\infty}{\cal C}_{min}^{asym}({\bf G})=1+\log_2\frac{\pi}{e}\simeq1.2088
$ (dot line in Fig.~\ref{fig1}).
This value is the asymptotic communication complexity of a noiseless quantum channel with the
constraint that the quantum states and the eigenstates of the measurements correspond to
Bloch vectors lying on a plane. In Ref.~\cite{montina}, we found a protocol for any quantum
state and projective measurements with communication cost equal to $\log_2(4/\sqrt{e})\simeq1.2786$,
which is about $6\%$ higher (dashed line in Fig.~\ref{fig1}). 
It is not known if this value is actually the asymptotic 
communication complexity of the quantum channel for general projective measurements.

In conclusion, we have presented a general procedure for evaluating the communication
complexity of channels in any general probabilistic theory, in particular quantum 
theory. This procedure, which relies on the reverse Shannon theorem and a strategy 
introduced in Refs.~\cite{montina,montina3}, is constructive and provides a method
to derive the most efficient protocol that classically simulates a channel. More
explicitly, given a quantum channel, we have defined a set ${\cal V}$ of 
classical channels and proved that the minimal classical capacity in $\cal V$ is the
asymptotic communication complexity of the quantum channel. Thus, the problem
of evaluating the communication complexity is reduced to a minimax problem. The 
channel in $\cal V$ with minimal capacity can be turned into the most efficient 
classical protocol for simulating the quantum channel. We have illustrated this 
procedure by evaluating the asymptotic communication complexity of a binary quantum 
depolarizing channel for some finite for sets of quantum states and measurements. 
The procedure is numerically very stable, but the computational time of the minimax
routine can grow exponentially with the number of quantum states and measurements.
Thus, specific strategies reducing the computational complexity need to be devised 
in the case of a high number of states and measurements. 

At the present, it not known if the communication complexity of noiseless quantum
channels is finite, unless the quantum channel capacity is $1$ qubit. Our method
can help to solve this open problem and, furthermore, to construct explicit
simulation protocols. As discussed in Ref.~\cite{montina},
the existence of finite classical communication protocols
is also deeply related to the existence of $\psi$-epistemic theories, which
are being object of recent intense study.

{\it Acknowledgments.} 
This work is supported by the Swiss National Science Foundation, 
the NCCR QSIT, and the COST action on Fundamental Problems in Quantum Physics.
A. M. acknowledges a support in part from Perimeter Institute for Theoretical
Physics, where a sketch of the proof of Theorem~1 was conceived.
Research at Perimeter Institute for Theoretical Physics is
supported in part by the Government of Canada through NSERC
and by the Province of Ontario through MRI.

\appendix
\section{Appendix}

{\it Proof of Theorem 1}.
Lemma 1 implies that ${\cal C}_{min}^{asym}({\bf G})\le {\cal D}({\bf G})$. 
We show that ${\cal C}_{min}^{asym}({\bf G})$ is actually equal to ${\cal D}({\bf G})$ by
proving that the asymptotic communication cost cannot be smaller than ${\cal D}({\bf G})$.
Let ${\cal C}_0$ be the asymptotic communication cost of a parallel simulation of the game $\bf G$.
We denote by $N$ the number of games $\bf G$ that are simulated in parallel. In the simulation,
Alice sends a variable $k$ generated
with conditional probability $\rho(k|y,s^1,\dots,s^N)$, where $s^i$ is an index labelling
the quantum state of the $i$-th game (hereafter superscripts label the game).
Bob simulates the measurements ${\cal M}_{m^1},\dots,{\cal M}_{m^N}$
by generating the outcomes $w^1,\dots,w^N$ according to a conditional probability 
$P(w^1,\dots,w^N|k,y,m^1,\dots,m^N)$. Let us denote by 
$P^i(w^i|k,y,m^1,\dots,m^N)$ the marginal probability of the outcome of
the $i$-th game.
We introduce the conditional probabilities
\be\begin{array}{c}
P^i(w_1,w_2,\dots,w_M|k,y)\equiv  \\
\prod_{m=1}^M P^i(w_m|k,y,1,1,\dots,{m^i}=m,\dots,1),
\end{array}
\ee
We will concisely denote $P^i(w_1,w_2,\dots,w_M|k,y)$ by $P^i({\bf w}|k,y)$.
Note that we have multiplied over the index $m^i$ and
set the other indices equal to $1$. For our purposes, any other choice of the values of the $N-1$ 
indices would be fine.
We use $P^i({\bf w}|k,y)$ to build the conditional probability
\be
P({\bf w}^1,\dots,{\bf w}^N|k,y)=\prod_i P^i({\bf w}^i|k,y).
\ee
Finally, from this distribution and $\rho(k|y,s^1,\dots,s^N)$, we build the conditional probability
\be\begin{array}{c}
\rho({\bf w}^1,\dots,{\bf w}^N|s^1,\dots,s^N)= \vspace{1mm} \\
\sum_k\int dy\rho(y)P({\bf w}^1,\dots,{\bf w}^N|k,y)\rho(k|y,s^1,\dots,s^N).
\end{array}
\ee
From the data-processing inequality~\cite{cover},
we have that the capacity, say 
$C({\bf W}^1,\dots,{\bf W}^N|S^1,\dots,S^N)$, 
of $\rho({\bf w}^1,\dots,{\bf w}^N|s^1,\dots,s^N)$ is smaller than or equal to the communication 
cost $N\, {\cal C}_0+o(N)$, that is, 
\be\label{first_in}
C({\bf W}^1,\dots,{\bf W}^N|S^1,\dots,S^N)\le N\, {\cal C}_0+o(N).
\ee
By construction, we have the constraints
\be\label{constr2}
\sum_{{\bf w}^1,\dots,{\bf w}^N,w^i_m=w}\rho({\bf w}^1,\dots,{\bf w}^N|s^1,\dots,s^N)=
P_{\bf G}(w|s_i,m),
\ee
the left-hand side being the marginal distribution of the variable $w_m^i$ (renamed $w$) 
given $s^1,\dots,s^N$.
Let $\rho_0({\bf w}|s)$ be the probability distribution in ${\cal V}({\bf G})$ with minimal capacity 
${\cal D}({\bf G})$. Then, it is easy to realized that the probability distribution
\be
\rho_{min}({\bf w}^1,\dots,{\bf w}^N|s^1,\dots,s^N)\equiv\prod_i\rho_0({\bf w}^i|s^i),
\ee 
is the channel satisfying constraints~(\ref{constr2}) with minimal capacity.
The minimum is equal to $N\,{\cal D}({\bf G})$. Thus, 
\be
N\,{\cal D}({\bf G})\le C({\bf W}^1,\dots,{\bf W}^N|S^1,\dots,S^N).
\ee
From this inequality and Inequality~(\ref{first_in}) we have that
\be
N {\cal D}({\bf G})\le N {\cal C}_0+o(N).
\ee
The theorem is proved. $\square$

\bibliography{biblio.bib}

\end{document}